# Comment on: "Dynamics of glass-forming liquids. XIII. Microwave heating in slow motion'' [J. Chem. Phys. 130, 194509 (2009)]

R. M. PICK

*IMPMC, Université P. et M. Curie et CNRS-UMR 7590, Paris France,*

*We show that the time dependent Dielectric Hole Burning experiments performed on supercooled liquids in the commented paper should be analyzed with a much shorter time step. The results would then exhibit strong short times oscillations, completely ignored in that paper. Also, different data corresponding to the same liquid are inconsistent ones with the others. They impede comparison between theory and experiment.*

In a recent paper [1] W. Huang and R. Richert summarized some recent Dielectric Hole Burning (D H B) experiments (references therein) performed in Dr Richert's group on supercooled liquids close to their glass transition temperature, $T_g$. They also proposed a full theoretical description of those experiments and performed some new, time resolved, ones that would simultaneously test the validity of the approach and reveal new aspects.

The basic D H B idea is that those liquids are an assembly of independent Dynamical Heterogeneities (D H), each of them having a Debye-like dynamics with relaxation time, $\tau$, $G(\tau)$ being the corresponding distribution function. When subjected to a (strong) electric field with angular frequency $\omega$, each D H independently absorbs energy due to its own $\tau$, is thus heated, and must be characterized, at time t, by the change of its local temperature, $\Delta T(\tau, t)$. This $\Delta T(\tau, t)$ changes, in turn, $\tau$ by $\Delta \tau(t)$, which modifies the dielectric function, $\varepsilon(\omega)$, $G(\tau)$ remaining unchanged.

Ref [1] gathered some older results related to this approach, and tested the preceding idea in detail on two molecular glass forming liquids, MTHF at T= 96.1 K and PC at T=166 K, temperatures chosen to correspond approximately to T=1.1 $T_g$. We discuss here two aspects of [1]. One is the meaning, at short times, of the recorded time dependent data. The second is the consistency of some of the presented results.

The method used to obtain those data is the following. Each D H is subjected to an electric field

$$E(t) = E_1 \sin(\omega t) \qquad t>0 \qquad (1)$$

and one extracts information from the time dependence of the part of the polarization of the sample which is cubic in $E_1$ and varies at angular frequency $\omega$. This is done, here as in [I], through a technique proposed in [2]: for a D H with relaxation time $\tau$, its polarization, P [3], is related to the instantaneous value of the electric field, E, by:

$$\tau\dot{P} + P = \varepsilon_0\chi E, \qquad (2)$$

$\chi$ being its susceptibility. If $\tau$ is weakly time dependent, Eq. 2 can be solved by a perturbation technique in $E_1$ in which $\tau$ is changed into $\tau + \Delta\tau$ and P into $P_1 + P_3$, $\Delta\tau$ (resp. $P_3$) being of order $E_1^2$ (resp. $E_1^3$):

$$\tau\dot{P}_1 + P_1 = \varepsilon_0\chi E, \qquad (2a)$$

$$\tau\dot{P}_3 + P_3 = -\frac{\Delta\tau}{\tau}\,\tau\,\dot{P}_1. \qquad (2b)$$

$\Delta\tau/\tau$ is caused by the heating, $\Delta T(\tau,t)$, of the D H and, in the hypothesis of a transfer of the D H extra energy to the heat bath governed by the same relaxation time $\tau$, its energy balance reads:

$$\Delta C_p\left(\Delta\dot{T} + \Delta T\big/\tau\right) = \frac{\tau\,\dot{P}_1^2}{\varepsilon_0\chi}, \qquad (3)$$

the r. h. s. of Eq. 3 being the instantaneous power not stored as an electromagnetic energy, $\Delta C_p$ being the D H heat capacity. $\Delta T$ is thus proportional to $E_1^2$ and so is $\Delta\tau/\tau$ where

$$\frac{\Delta\tau}{\tau} = -\frac{\Delta T}{A}. \qquad (4)$$

One can solve analytically, step by step, Eqs. 2a, 3, and 2b with E given by Eq. 1 and the corresponding results are given in the Appendix. $P_3^{tot}(t)$, the total polarization of the liquid is finally obtained by summing $P_3$ over the distribution function. The changes in the real and imaginary parts of $\Delta\varepsilon(\omega,t)$ are then assumed, following [1], to be given by

$$\Delta\varepsilon'(\omega,t)\ (resp.\ \Delta\varepsilon''(\omega,t)) = \frac{2}{\varepsilon_0 T}\frac{1}{E_1}\int_{t-kT}^{t+(1-k)T}\sin\omega t'\ (resp.\ (-)\cos\omega t')\ P_3^{tot}(t')\ dt', \qquad (5)$$

with $\omega T = 2\pi$ and $0 < k < 1$. In the spirit of the perturbation approach used here, the results, presented, e. g. in Figs. 6 to 9 of [1], can be computed from:

$$\Delta\log\,\mathrm{tg}(\delta(t)) = \frac{\Delta\varepsilon''(\omega,t)}{\varepsilon''(\omega)} - \frac{\Delta\varepsilon'(\omega,t)}{\varepsilon'(\omega)}, \qquad (6)$$

where $\varepsilon'(\omega)$ (resp. $\varepsilon''(\omega)$) are the corresponding equilibrium values [4].

The continuous lines in Fig. 1 are the results, for MTHF at 96.1 K, of such a calculation for $\nu = \omega/2\pi = 500$, 1000, and 2000 Hz, using the dielectric function proposed in [1], k being equal to unity, as it is presumably the case in [1]. The initial jump at t=T is a mathematical artifact, consequence of the integration of $P_3^{tot}(t)$ over one period, Eq. 5: $\Delta\varepsilon(\omega,t)$ cannot be defined for t<T, and it has an unclear physical meaning in its immediate vicinity. The strong short times oscillations of this figure have a different origin, namely that $P_3^{tot}(t)$, Fig. 2, can be approximated, for every value of $\omega$ considered in [1], by

$$P_3^{tot}(t) = f(t)\sin(\omega(t - t_0))$$  (7)

with $t_0 \approx 0.17T$ and $f(t=0) = 0$: $f(t)\sin(\omega t)\sin(\omega(t - t_0))$ has thus a $f(t)\cos(\omega(2t - t_0))$ component. As $f(t)$ is not a constant, $\Delta\varepsilon'(\omega,t)$ exhibits oscillations with a period 0.5 T, and the same is true for $\Delta\varepsilon''(\omega,t)$.

Neither the initial jump at t=T, nor the preceding oscillations appear when one uses the "mean power approximation" in which, see [1, 6, and 7], the r. h. s. of Eq. 3 is replaced, by its mean value for t→∞, and the corresponding $\Delta T(\tau,t)$ is inserted in Eq. 4. $\Delta\varepsilon(\omega,t)$ is then obtained from

$$\Delta\varepsilon'(\omega,\tau,t) \ (resp. \ \Delta\varepsilon''(\omega,\tau,t)) = -K\frac{a}{1 + a^2}\text{Im} \ (resp. \ \text{Re})\left[a\frac{d}{da}\left(\frac{1}{1 + ia}\right)\frac{\Delta\tau(t)}{\tau}\right] \quad a = \omega\tau,$$  (8)

summing the l. h. s. of Eq. 8 over $G(\tau)$, with K independent of $\tau$. The three dashed lines of Fig. 1 result from that approximation for the same three frequencies, the results being rescaled at t=8T for each frequency: as mentioned in [1], for the same frequency, the two curves do not coincide for t<4T.

Unfortunately, all the figures contained in [1] which correspond to the continuous lines of Fig.1 present the experimental results only for discrete time values, t= (n+α)T, n being a positive integer and 0< α<1, while the value of α, never given, varies, on the same figure, with the measurement frequency ν. These figures also contain, when presented on a linear time scale, a straight line between an assumed 0 value for t=0 and the t=(1+α)T value of $\Delta\log tg(\delta(t))$. The slope of this line seems to be the quantity reported versus ν in the inset of Fig. 12 of [1]. This slope has no clear physical meaning because of its large dependence on the value of α (which varies with ν) and of the artificial value of $\Delta\varepsilon(\omega,t)$ in the vicinity of t=T, as mentioned above. Consequently, the same is true for the curvature of the curve represented by those two points and the t=(2+α)T one. Conclusions based on the "short times" values of $\Delta\log tg(\delta(t))$, as presented in [1] and [7], are thus doubtful. They need to be strengthened by measurements performed with a much shorter time interval in the T< t<2.5 T range.

The theoretical results of Fig.1 unfortunately cannot be compared with the MTHF data presented on Figs. 6 and 7 of [1], for the same temperature and frequencies, for $E_1$=283 KV/cm, during a eight periods heating experiment. Indeed, those results are inconsistent between the two figures when both plotted versus t/T, as done here on Fig. 3: the 500 and 1000 Hz data points of Fig. 6 are below those of Fig. 7 while it is the opposite in the 2000 Hz case. Conclusions, based on the results presented on Fig. 7 of [1], for those three and for higher frequencies, are thus doubtful, while this figure is the basis of the discussions performed in the last but one paragraph of [1] (Relation to Physical Aging), and in part of [7].

Other results, presented in [1], Figs. 5, 8, or 9, for PC, also need to be reconsidered. These results have been obtained with three different values of $E_1$ for the three figures. The $\Delta \log \mathrm{tg}(\delta(t))$ measurements, for a pump frequency $\nu=1000$ Hz, presented on Figs. 8, and 9 are consistent: they scale by the $E_1^2$ factor. Conversely, they are not consistent with the results shown on Fig. 5c. The latter represents the long-time value of $\Delta \log \mathrm{tg}(\delta(t))$ for many values of $\nu$. For $\nu= 1000$ Hz, this value can also be obtained from Fig. 9 for $t \approx 60$ ms. The two values do not scale with $E_1^2$: the first one is too large by a factor of $\approx 2$. Also, and more important, Figs. 5a and 5b. are inconsistent one with the other. As explained in [1], one can extract from the value of $\Delta \varepsilon''(\omega, t \to \infty)$, Fig. 5a of [1], and from the analytical expression of $\varepsilon(\omega)$, the opposite of the relative change in frequency, $\Delta \nu / \nu$, which would produce the same $\Delta \varepsilon''(\omega)$ when keeping the temperature fixed. We have performed this computation, using the data points of this Fig. 5a and the analytical form of the dielectric function given in [1]. Those $\Delta \nu / \nu$ values do not agree with the data points reported on Fig. 5b over the entire $\nu > 20$ Hz range, as shown on Fig. 4. As for MTHF, it is thus not possible to decide which PC data are the correct ones.

In conclusion, though the method proposed in [1] is interesting and, for instance, the possible t/T scaling in those DHB experiments for angular frequencies 25 to 100 times the frequency of the maximum of $\varepsilon''(\omega)$ is an important finding, much more experimental results are necessary to ascertain their internal consistency as well as their consistency with the proposed theoretical model. Also, a more physically meaningful definition of $\Delta \varepsilon(\omega, t)$ needs to be obtained to unravel the physical content of those experiments at short times.

## Figures caption

FIG. 1: $\Delta \tan(\delta)/\tan(\delta)$ versus t/T : 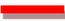 500 Hz, 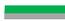 1000 Hz, 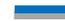 2000 Hz. Continuous lines: extracted from $P_3^{tot}(t)$, Eq. 5. Dash and dot lines: computed from the "mean power approximation" (see Eq. 8 and the related text).

FIG 2: $P_3^{tot}(t)$ versus t/T : 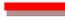 500 Hz, 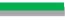 1000 Hz, 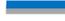 2000 Hz.

FIG 3: $\Delta \tan(\delta)/\tan(\delta)$, following [1], versus time.  500 Hz,  1000 Hz,  2000 Hz, extracted from Fig. 7a;  500Hz,  1000 Hz,  2000 Hz, extracted from Fig. 6.

FIG 4: , $\Delta \varepsilon''(\omega)/\varepsilon''(\omega)$ following [1], Fig. 5a; , $\Delta \nu / \nu$ following [1], Fig. 5b; , $\Delta \nu / \nu$, our calculation, using the $\Delta \varepsilon''(\omega)/\varepsilon''(\omega)$ data from Fig. 5a of [1] and the analytical expression of $\varepsilon(\omega)$ (Eq. 14 of [1]). The inclusion of the second term of Eq. 14 does not change the result by more than 2 %.

## Appendix

We give here the results for the successive integrations of Eqs. 2a, 3, and 2b, making use of Eq. 4 to relate $\Delta\tau$ and $\Delta T$ and of the expression of $E(t)$ given in Eq. 1. In order to simplify the results the notations $u = \dfrac{t}{\tau}$ and $a = \omega\tau$ are introduced as well as the amplitudes, $A_n$, and the phases, $p_n$. The latter are related to the variable $a$ by

$$A_n = \sqrt{1+(na)^2} \ , \quad \cos p_n = \frac{1}{A_n}, \quad \sin p_n = \frac{na}{A_n}$$

In terms of those notations, the solutions for the three equations read:

<u>Eq.2a</u>:
$$P_1(t) = \frac{\varepsilon_0 \chi\, E_1}{A_1}\left[\sin(\omega t - p_1) + \sin p_1 \ \exp(-u)\right].$$

<u>Eq. 3</u>:
$$\Delta T(t) = \frac{\varepsilon_0 \chi\, E_1^2}{\Delta C_p}\left(\frac{\sin p_1}{A_1}\right)^2 (K_1 + K_2) \quad \text{with}$$

$$K_1 = \frac{A_1^2}{2}\left[(1-\exp(-u)) + \frac{1}{A_2}\left(\cos(2\omega t - 2p_1 - p_2) - \exp(-u)\cos(2p_1 + p_2)\right)\right],$$

$$K_2 = -\exp(-u)\left[1 + \exp(-u) + 2\frac{\sin(\omega t - p_1)}{\sin p_1}\right].$$

<u>Eq. 2b</u>
$$P_2(t) = \frac{(\chi\varepsilon_0)^2 E_1^3}{A\ \Delta C_p}\left(\frac{\sin p_1}{A_1}\right)^3\left[\frac{\exp(-u)}{2}\left[L_1 + L_2 + L_3 + L_4\right] + M_1\right] \quad \text{with}$$

$$L_1 = -\left[A_1^2(u + \cos 2p_1) + \frac{1}{2}\frac{A_1^3}{A_2}\left(\frac{\cos(3p_1 + p_2 + p_3)}{A_3} + \frac{\sin(3p_1 + p_2)}{a}\right) + 1 + \frac{\cos(2p_1)}{(\sin p_1)^2}\right],$$

$$L_2 = -\left[\left(1 + A_1^2 - \left(\frac{3a}{A_2}\right)^2\right)\left(\exp(-u) + \frac{\sin(\omega t - p_1)}{\sin p_1}\right) + \exp(-2u)\right],$$

$$L_3 = \frac{A_1^2}{A_2^2}\left[(1 + A_2^2)\cos(2\omega t - 2\,p_1) - \frac{1}{2a}\sin(2\omega t - 2\,p_1)\right],$$

$$L_4 = -\exp(-u)\frac{4\,A_{0.5}}{a}\sin(\omega t - p_{0.5}),$$

$$M_1 = \frac{A_1^2}{2} \left[ \cos(\omega t - 2\,p_1) + \frac{1}{2\,A_2} \cos(\omega t - 2\,p_1 - p_2) + \frac{A_1}{2\,A_2\,A_3} \cos(3\omega t - 3\,p_1 - p_2 - p_3) \right].$$

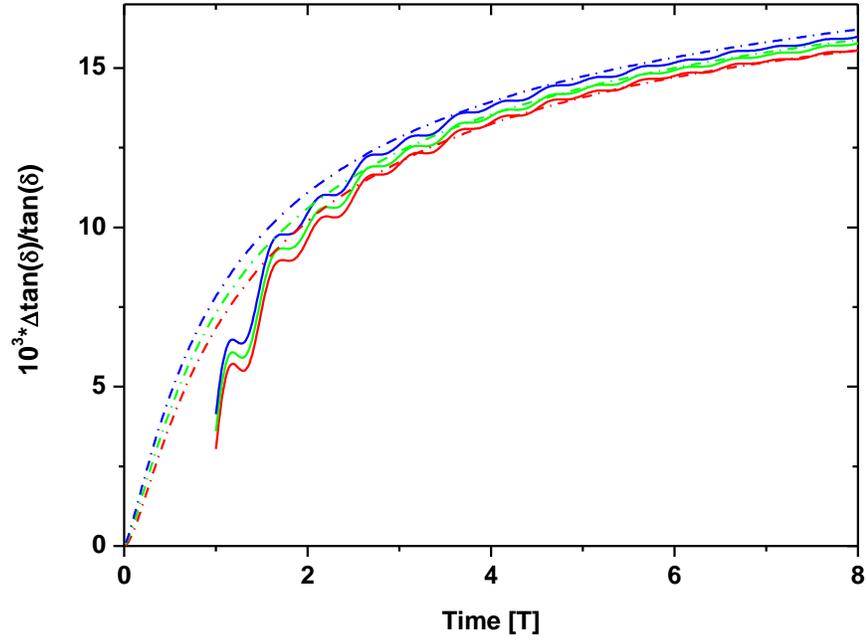

Fig .1

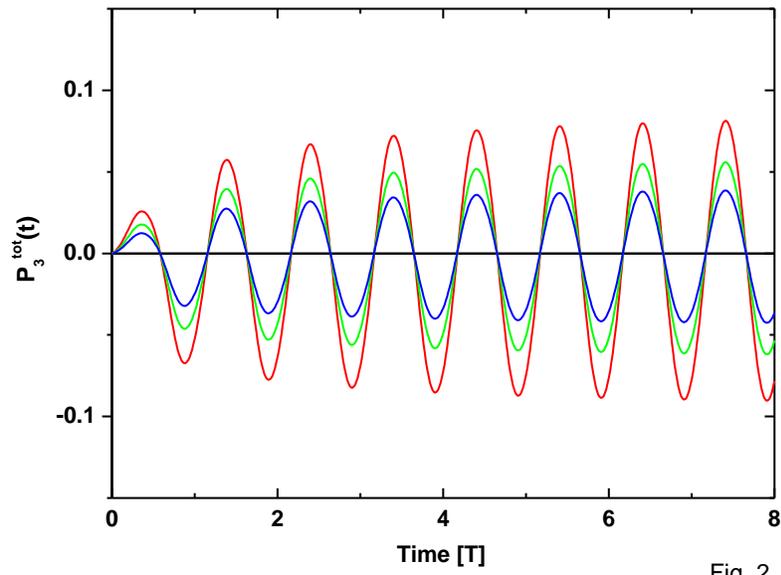

Fig. 2

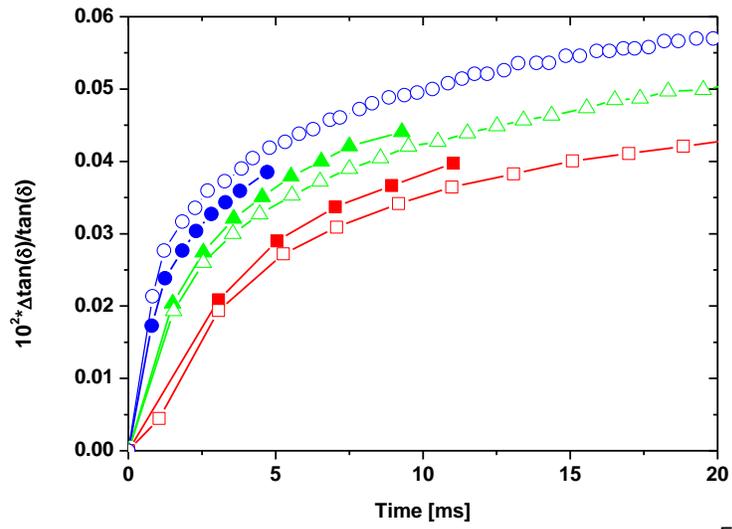

Fig. 3

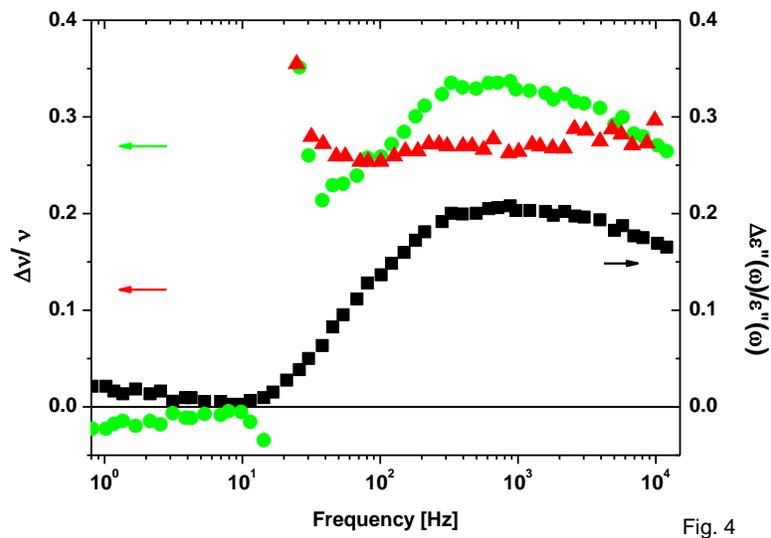

Fig. 4